\documentstyle[pre,aps,epsf,twocolumn]{revtex}
\begin{document}
\title{Conformal Theory of the Dimensions of Diffusion Limited Aggregates}
\author{Benny Davidovich and Itamar Procaccia}
\address{Dept. of Chemical Physics, The Weizmann Institute of Science, Rehovot
76100, Israel}
\maketitle
\begin{abstract}
We employ the recently introduced conformal iterative construction of
Diffusion Limited Aggregates (DLA) to study the multifractal
properties of the harmonic measure. The support of the harmonic measure is 
obtained from a dynamical process which is complementary
to the iterative cluster growth. We use this method to
establish the existence of a series of random scaling functions that yield, 
via the thermodynamic formalism of multifractals, the generalized dimensions 
$D_q$ of DLA  for $q\ge 1$. 
The scaling function is determined just by the last stages of the iterative 
growth process which are relevant to the complementary dynamics. Using the
scaling relation $D_3=D_0/2$ we estimate the fractal dimension of DLA
to be $D_0=1.69\pm 0.03$. 
\end{abstract}
\vskip 0.3 cm
The diffusion limited aggregation (DLA) model was introduced in 1981 by 
T. Witten and L. Sander \cite{81WS}. DLA has attracted enormous amount
of research as an elegant example of the dynamical creation of a non-trivial
fractal set, and as a model that underlies many
pattern forming processes including dielectric breakdown \cite{84NPW}, two-fluid
flow \cite{84Pat}, and electro-chemical deposition \cite{89BMT}. In spite
of the significant amount of effort to understand the fractal and the
multifractal properties of DLA, there exists to date no accepted calculation 
of these properties from first principles. In this Letter we present a theory
that is based on an iterative conformal construction of DLA \cite{98HL,98DHOPSS}
that culminates with the construction of a series of (random) scaling functions which allow, via the
thermodynamic formalism of multifractals \cite{83HP,86HJKPS,88Mea}, a convergent calculation 
of the multifractal properties of DLA.

Originally the DLA model was introduced as the outcome of N random walks.
Fixing one particle at the center of coordinates in $d$-dimensions, and 
releasing random walkers from infinity one at a time, one allows them to walk around 
until they hit any particle belonging to the cluster. Upon hitting they are attached to the
growing cluster. We are interested here in $d=2$ for which numerical simulations 
indicated that for $N\to \infty$ the cluster attains a fractal dimension of about 
1.71 \cite{88Mea}.  The approach used 
here is different: we employ an iterative conformal construction of DLA that was recently
proposed by Hastings and Levitov \cite{98HL}. The basic idea is to follow the evolution of 
the conformal mapping $\Phi^{(n)}(w)$ 
which maps the exterior of the unit circle in the mathematical $w$--plane onto 
the complement of the cluster of $n$ particles in the physical $z$--plane.
The unit circle is mapped to the boundary of the cluster
which is parametrized by the arc length $s$, $z(s)=\Phi^{(n)}(e^{i\theta})$.
This map $\Phi^{(n)}(w)$ is made from compositions of elementary maps $\phi_{\lambda,\theta}$,
\begin{equation}
\Phi^{(n)}(w) = \Phi^{(n-1)}(\phi_{\lambda_{n},\theta_{n}}(w)) \ , \label{recurs}
\end{equation}
where the elementary map $\phi_{\lambda,\theta}$ transforms the unit circle
to a circle with a ``bump" of linear size $\sqrt{\lambda}$ around the point
$w=e^{i\theta}$. Accordingly the map $\Phi^{(n)}(w)$ adds on a new bump to the
image of the unit circle under $\Phi^{(n-1)}(w)$. The bumps in the physical
$z$-plane simulate the accreted random walkers in the original formulation. 
The main idea in this
construction is to choose the positions of the bumps $\theta_n$ and their
sizes $\sqrt{\lambda_n}$ such as to achieve accretion of {\em fixed linear size}
bumps on the boundary of the growing cluster according to the harmonic measure
$P(s)ds$. The latter is the probability that a random walker would hit
an infinitesimal arc $ds$ centered at the point $z(s)$. 
This is done as follows.

The probability density $P(s)$ is given by the inverse of the derivative of the conformal 
map $P(s)= \left[{\Phi'}^{(n)}(e^{i\theta})\right]^{-1}$. Using the obvious fact
that $ds=|{\Phi'}^{(n)}(e^{i\theta})| d\theta$ we see that 
$P(s)ds\equiv d\theta$. i.e. the harmonic measure is uniform on the unit circle.
Thus choosing random positions $\theta_n$, and $\lambda_n$ in Eq.(\ref{recurs}) according to
\begin{equation}
\lambda_{n} = \frac{\lambda_0}{|{\Phi^{(n-1)}}' (e^{i \theta_n})|^2} \label{lambdan}
\end{equation}
we accrete fixed size bumps in the physical plane according to the harmonic measure.
Finally the elementary map $\phi_{\lambda,\theta}$ is chosen as
\begin{eqnarray}
&&\phi_{\lambda,0}(w) = w^{1-a} \left\{ \frac{(1+ \lambda)}{2w}(1+w)\right. 
\nonumber\\
&&\left.\times \left [ 1+w+w \left( 1+\frac{1}{w^2} -\frac{2}{w} \frac{1- \lambda}
{1+ \lambda} \right) ^{1/2} \right] -1 \right \} ^a \\ &&\phi_{\lambda,\theta} (w) 
= e^{i \theta} \phi_{\lambda,0}(e^{-i \theta} w) \,, \label{eq-f}
\end{eqnarray}
The parameter $a$ is confined in the range $0< a < 1$, determining the shape
of the bump. In this Letter we employ $a=2/3$ which is consistent with
semicircular bumps. The qualitative properties of this mapping that enter
prominently our analysis are the following:
\begin{equation}
\phi_{\lambda,\theta}(w)\approx (1+\lambda)^a w\quad 
{\rm for}~ {|w|-1\over \sqrt{\lambda}}\ge {\rm const} \ , \label{phidis}
\end{equation}
where const here is a tolerance-dependent but $\lambda$-independent constant
of the order of unity.
\begin{eqnarray}
|\phi_{\lambda,\theta}(w)|&\approx& (1+\sqrt{\lambda})|w|
\quad{\rm for~ arg} w\in[\theta\pm\sqrt{\lambda}]\ , \label{distort}\\
|\phi_{\lambda,\theta}(w)|&\approx& |w|
\quad{\rm for~ arg}w\not\in[\theta\pm\sqrt{\lambda}] \ . \label{reparam}
\end{eqnarray}
Eq.(\ref{reparam}) means that points that do not belong to the bump are
only reparametrized along the circle.

The recursive dynamics can be represented as iterations of the map 
$\phi_{\lambda_{n},\theta_{n}}(w)$,
\begin{equation}
\Phi^{(n)}(w) =
\phi_{\lambda_1,\theta_{1}}\circ\phi_{\lambda_2,\theta_{2}}\circ\dots\circ 
\phi_{\lambda_n,\theta_{n}}(\omega)\ . \label{comp} 
\end{equation}
The final calculation of $\lambda_n$ can be done many times (for 
a given history $\theta_1,\theta_2,\dots,\theta_{n-1}$) according to the
uniform measure on the unit circle, yielding a distribution of values of
$\lambda_n$, which according to Eq.(\ref{lambdan}) is the distribution
of the density of the harmonic measure. The moments of this distribution
are connected to the generalized dimensions of the harmonic measure.
The latter are defined as follows \cite{86HJKPS}: consider a partition of the cluster boundary
into balls of diameter $\ell_i$ and measure $p_i$. The generalized
dimensions $D_q$ are determined by the equation
\begin{equation}
\lim_{M\to \infty} {\sum_{i=1}^M {p_i^q\over \ell_i^{(q-1)D_q}}} =1 \ . \label{HP}
\end{equation}
The calculation of $D_q$ from the statistics of $\lambda_n$ was discussed in 
detail in Ref.\cite{98DHOPSS} with the exact result
\begin{equation}
\overline {\lambda^q_n} \sim n^{-2qD_{2q+1}/D} \ . \label{lamDq}
\end{equation}
It is well known \cite{86HMP} that the fractal dimension is $D_0$ in this language,
the information dimension (known also as the dimension of the harmonic measure) 
is $D_1=1$ \cite{85Mak}, and in general
$D_q\ge D_{q'}$ for any $q'>q$. It was shown in \cite{87Hal} and in 
\cite{98DHOPSS} that $D_3=D_0/2$. This last result means of course that
on the average $\lambda_n$ decreases like $n^{-1}$. The exact result
found in \cite{98HL,98DHOPSS} is 
\begin{equation}
\overline \lambda_n =\frac{1}{aDn} \ . \label{avlam}
\end{equation}

Consider the support of the harmonic measure. We use the uniformity
of the measure on the unit circle to form, for a given cluster
of $n$ bumps, an equi-measure partition
of $M$ balls in the physical space, by selecting points $z_{k}$
\begin{equation}
z_{k}\equiv \Phi^{(n)}(e^{i\theta_k}) =
\phi_{\lambda_1,\theta_{1}}\circ\phi_{\lambda_2,\theta_{2}}\circ\dots\circ 
\phi_{\lambda_n,\theta_{n}}(e^{i\zeta_k})\ ,
\end{equation}
where the points $\zeta_k$, $k=1,2,\dots,M$ are uniformly spaced on the unit circle. 
We introduce the ``complementary dynamics" to the cluster
growth by
\begin{equation}
z_{j,k}\equiv 
\phi_{\lambda_{j},\theta_{j}}\circ\phi_{\lambda_{j+1},\theta_{j+1}}\circ\dots\circ 
\phi_{\lambda_n,\theta_{n}}(e^{i\zeta_k})\ ,
\end{equation}
and $z_{1,k}= z_k$.
The complementary dynamics creates the points $z_k$ from
the seeds $e^{i\zeta_k}$. In Fig.1 we display the typical evolution of the set 
$\{z_{j,k}\}$ for different values of $j$. The striking observation is that  the
equi-measure partition is fixed {\em in shape} very early (large $j$) in the complementary
dynamics. Later stages only serve to inflate the fixed shape, with
the last steps being most prominent in determining the final radius of
this set which of course is the radius of the cluster. 
\vskip 1 cm
\narrowtext
\begin{figure}
\epsfxsize=8.0truecm 
\epsfysize=8.0truecm
\epsfbox{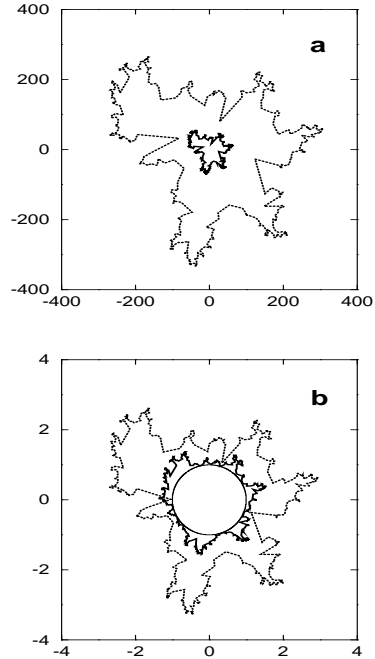}
\vskip 1 cm
\caption{The evolution of the set $\{z_{j,k}\}$. (a) $j$=11 and 1, 
(b) $j=10^4$, 6000 and 2000. Note that the straight lines are not parts
of the set, they simply connect points on the set.}
\label{dragons}
\end{figure}
This observation is easy to understand. As noted in Eqs.(\ref{phidis}-\ref{reparam}) 
the elementary map $\phi_{\lambda,\theta}(w)$
distorts circles of radius close to unity, but acts as a uniform multiplication
for points with larger absolute values. For $j=n$ we always start the complementary dynamics
on the unit circle, and (\ref{distort}-\ref{reparam}) are applicable. We
estimate now how many iterations of the complementary dynamics are necessary
before (\ref{phidis}) becomes applicable. Consider an arbitrary value
of $\zeta_k$. The first iteration of the complementary dynamics $z_{n,k}$  
have $|z_{n,k}|\approx 1+\sqrt{\lambda_n}$ iff $\zeta_k\in[\theta_n\pm\sqrt{\lambda_n}]$,
an event of probability $\kappa_n\equiv \sqrt{\lambda_n}/2\pi$. Otherwise $|z_{n,k}|=1$.
If the dynamics failed to increase $|z_{n,k}|$, it can do it with $|z_{n-1,k}|$
after two steps, with probability $(1-\kappa_n)\kappa_{n-1}$.
The average number of steps $A$ needed to grow out with certainty is therefore 
\begin{equation}
A=\sum_{k=0}^\infty k(1-\kappa_n)(1-\kappa_{n-1})\dots
(1-\kappa_{n-k+2})\kappa_{n-k+1}
\end{equation}
Averaging over the history $\lambda_1,\dots,\lambda_n$, approximating
$\langle\lambda_{n-k}\lambda_{n-k'}\rangle\sim \langle\lambda_{n-k}\rangle\langle\lambda_{n-k'}
\rangle$ (which
was justified for $k\ne k'$, $k,k'\ll n$ in \cite{98DHOPSS}), we find
\begin{equation}
\langle A\rangle=\sum_{k=0}^\infty k(1-\langle\kappa_n\rangle)\dots
(1-\langle\kappa_{n-k+2}\rangle)\langle\kappa_{n-k+1}\rangle
\end{equation}
We assume and show self-consistently that the sum is dominated by $k\ll n$.
In that case 
\begin{equation}
\langle A\rangle\approx\sum_{k=0}^\infty k(1-\langle\kappa_n\rangle)^{k-1}\langle\kappa_{n}\rangle
=\frac{1}{\langle\kappa_n\rangle}\sim n^{D_2/D} \ , \label{wow}
\end{equation}
where (\ref{lamDq}) has been used. The variance is estimated similarly, and is of the
same order. Since $0.5=D_3/D \le D_2/D \le D_1/D=0.58$ we
see that for large $n$ we need relatively few iterations of the complementary
dynamics to increase the radius of the set $\{z_{j,k}\}$ to a value after which
the overwhelming majority of the iterations serve simply to inflate the
radius by factors of $(1+\lambda_j)^a$. In light of Eq.(\ref{avlam}) we understand
the phenomenon exhibited in Fig.1, that the last steps of the
complementary dynamics have the largest effect on the radius of the set $\{z_{j,k}\}$.
We note in passing that these comments also explain the finding of \cite{98HL,98DHOPSS} that the 
first Laurent coefficient of $\Phi^{(n)}(w)$, denoted there as $F^{(n)}_1$
is a measure of the radius of the cluster. The exact result is that  $F^{(n)}_1=
\prod_{i=1}^n (1+\lambda_i)^a$. Finally note that the set $\{z_{j,k}\}$ for $k\to \infty$
generates the support of the harmonic measure. It will reveal the generalized
dimensions $D_q$ for $q\ge 1$ only.
\narrowtext
\begin{figure}
\epsfxsize=8.0truecm
\epsfysize=8.0truecm
\epsfbox{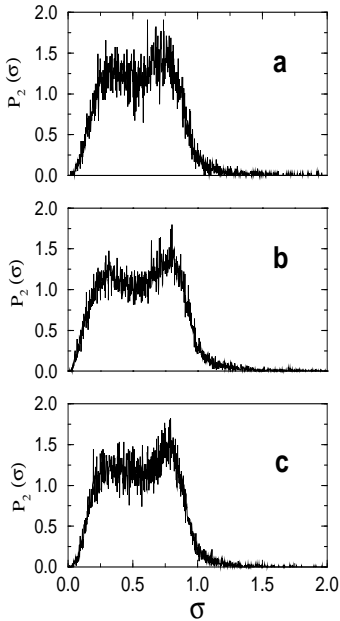}
\caption{(a) The random scaling function $P_2(\sigma)$ obtained from two consecutive 
partitions of 16 and 32 equi-measure balls. 550 clusters of
10000 bumps were used. (b) $P_2(\sigma)$ computed from 32 and 64 balls,
with 550 clusters of 10000 bumps. (c) $P_2(\sigma)$ computed from 32 and 64 balls,
with 44 clusters of 100 000 bumps.}
\label{scalefunc}
\end{figure}
At this point we can introduce a ``binary" scaling function. Consider two values of $M$,
$M_1=2^m$ and $M_2=2^{m+1}$. Consider the two associated sets $\{\tilde z_i\}_{i=1}^{M_1}$,
$\{z_k\}_{k=1}^{M_2}$ such that $\tilde z_i=z_{2i}$. For every difference 
$\tilde z_i-\tilde z_{i-1}$ in the coarser
resolution $M_1$ (denoted as the $ith$ ``mother"), there are two
differences $z_{2i}-z_{2i-1}$ and $z_{2i-1}-z_{2i-2}$ in the finer resolution,
which are denoted as ``daughters". The binary scaling function at every resolution 
has $M_2$ values which are obtained as the ratio of daughters to mothers, 
\begin{eqnarray}
\sigma_{2i}&\equiv& (|z_{2i}-z_{2i-1}|)/(|\tilde z_i-\tilde z_{i-1}|)\nonumber\\
\sigma_{2i-1}&\equiv& (|z_{2i-1}-z_{2i-2}|)/(|\tilde z_i-\tilde z_{i-1}|) \ . \label{defsig}
\end{eqnarray}

{\bf Conjecture}: For $1\ll M_2 \ll n$ the binary scaling function (\ref{defsig}) converges 
in distribution to a universal function independent of $M_2$ and the history $\theta_1,
\dots \theta_n$. We denote the distribution as $P_2(\sigma)$.

In Fig.2 we show the numerical evidence for this conjecture. We employ clusters with 
$n=10^4$ and $n=10^5$, and various values of $M_2$. Note that $M_2\ll n$; increasing
$M_2$ beyond, say, 128 for $n=10^5$ results in exposing the ultra-violet cutoff of the
smooth bumps, yielding a spurious peak at $\sigma=1/2$. The harmonic measure is extremely 
concentrated near the tips of the
cluster, and any attempt to resolve the fjords leads to oversampling of the smooth
bumps on the tips. Note that the scaling function is defined
as a ratio, and the discussion above implies that it is sensitive to only $n^{D_2/D}$
last growth steps. The majority of the iterations of the fundamental map 
$\phi_{\lambda,\theta}$ are irrelevant, as they cancel in the ratio.

Ordering the $2^n$ diameters $\tilde z_j-\tilde z_{j-1}$ in a binary basis
$\ell(\epsilon_m\dots\epsilon_1)$, redefining the ratios $\sigma(\epsilon_m\dots\epsilon_1)$
accordingly, and using in Eq.(\ref{HP}) the fact that 
$p_i=2^{-m}$ for the $M_1$ balls, we derive the equation \cite{86FJP}
\begin{eqnarray}
\sum_{\epsilon_{m+1}\dots\epsilon_1} \sigma^{-\tau(q)}(\epsilon_{m+1}\dots\epsilon_1)
&&\ell^{-\tau(q)}(\epsilon_m\dots\epsilon_1)\nonumber\\&&=2^q \sum_{\epsilon_m\dots\epsilon_1} 
\ell^{-\tau(q)}(\epsilon_m\dots\epsilon_1)\ ,
\end{eqnarray}
where $\tau(q)\equiv (q-1)D_q$. Iterating, we find
\FL
\begin{eqnarray}
S_{m+1}&\equiv&\!\!\!\sum_{\epsilon_{m+1}\dots\epsilon_1}\!\!\!\!\! \sigma^{-\tau(q)}
(\epsilon_{m+1}\dots\epsilon_1)
\sigma^{-\tau(q)}(\epsilon_m\dots\epsilon_1)\dots \sigma^{-\tau(q)}(\epsilon_1)
\nonumber\\&=&2^{(m+1)q}\ . \label{motek}
\end{eqnarray}
To compute $D_1$ we notice that $\tau(1)=0$. For $\tau(q)\to 0^+$ all
the realizations of the products of random numbers $\sigma^{-\tau}$ are comparable.
Since the most probable product for $m\to \infty$ is $\exp[{(m+1)\overline{\ln\sigma}}]$, 
we can estimate the sum in (\ref{motek}) as $2^{(m+1)}\exp[{(m+1)\overline{\ln\sigma}}]$. Substituting
in (\ref{motek}) and taking the $m$th root and the log to base 2 gives
\begin{equation}
\lim_{q\to 1^+} \tau(q) =\frac{1-q}{\overline{\log_2\sigma}} \ . \label{wow2}
\end{equation}   
Using the scaling function $P_2(\sigma)$ we compute $\overline{\log_2\sigma}\approx -1\pm 0.01$,
yielding the expected result $D_1=1$. This is the first nontrivial calculation of a
generalized dimension in this approach.

For $q>1$ the approach using the most probable product of random numbers is not
applicable. We have $2^m$ realizations of products of $m$ random numbers, and
rare events are relevant. Moreover, one should notice that even though every
factor of $\sigma^{-\tau}$ is independent, the products in the sum are not,
as they have common factors. One can use the fact that the random products 
are organized on a binary tree to write an exact recursion relation
\begin{equation}
S_{m+1} ={\sigma(0)}^{-\tau(q)}(0) S_m^{(0)} +{\sigma(1)}^{-\tau(q)} S_m^{(1)} \ , \label{rec2}
\end{equation}
where $ S_m^{(\epsilon)}$ are two independent realizations of the sum of $2^m$ products, each of $m$ random variables. We failed to find the exact asymptotics of 
$S_m$ which is necessary for computing $\tau(q)$. We recognize however that
{\em our binary partition is in fact arbitrary}, and instead we can perform 
refinements into $k$ daughters at each step. Accordingly we will have $k$ scaling
ratios $\sigma(\epsilon)$ for each mother, with $\epsilon$ now taking on $k$
values $k=0,1,\dots k-1$. The random scaling function is now denoted as $P_k(\sigma)$;
its existence as a universal function for $k=2^m$ emanates in an obvious fashion from
the existence of $P_2(\sigma)$, and its existence for any $k$ can be demonstrated independently
as done above for $P_2(\sigma)$. The important point is that the asymptotics of Eq.(\ref{motek})
is computable in the limit $k\to \infty$:
\begin{equation}
S_{m+1} =\sum_{\epsilon_1=0}^{k-1}{\sigma(\epsilon_1)}^{-\tau(q)} S_m^{(\epsilon_1)} \ . \label{motek2}
\end{equation} 
In the limit of $k\to \infty$ this equation reads 
\begin{equation}
S_{m+1} \to k \overline{{\sigma(\epsilon_1)}^{-\tau(q)} S_m^{(\epsilon_1)}}
= k\overline{{\sigma}^{-\tau(q)}}\overline{S_m}
\end{equation}
where we have used the fact that ${\sigma(\epsilon_1)}^{-\tau(q)}$ is random, independent of 
the {\em consecutive} factors  ${\sigma(\epsilon_1,\epsilon_2)}^{-\tau(q)}\dots$ which
consist $S_m^{(\epsilon_1)}$. Asymptotically $S_{m+1} \to [k\overline{{\sigma}^{-\tau(q)}}]^{m+1}$,
and substituting in (\ref{motek}) we compute $\tau(q)$ from
\begin{equation}
\int P_k(\sigma)\sigma^{-\tilde\tau(q)} d\sigma = k^{q-1}\ , \quad
\lim_{k \to \infty}\tilde\tau(q)=\tau(q) \ . \label{final}
\end{equation}
\hskip -1 cm
\narrowtext
\begin{figure}
\epsfxsize=5.0truecm
\epsfysize=5.0truecm
\epsfbox{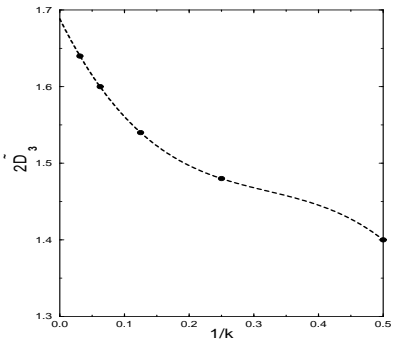}
\caption{$2\tilde D_3(k)$ as a function of $1/k$. We are interested in $D_0=2\tilde D_3(k\to \infty)$.
The best fit (dashed line) predicts $D_0$=1.69.}
\label{dimension}
\end{figure}
In Fig.3 we show the solution
$2\tilde D_3=\tilde\tau(3)$ from (\ref{final}) as a function of $1/k$. It should
be noted that when $k$ increases the determination of $\tilde\tau(q)$ becomes
easier since the LHS of (\ref{final}) changes from very small to very large
value over a narrow range of value of $\tilde\tau$. This is compatible with
the thermodynamic formalism of multifractals \cite{86FJP}. We expect the limit as $1/k\to 0$ to
be regular, and we fit a 3rd order polynomial to the points as shown in the figure.
The best fit predicts $2D_3=2\tilde D_3(k\to \infty)=1.69$. Examining different partitions of the unit
circle and different integration schemes we concluded that we can bound the errors 
around $D_0=2D_3=1.69\pm 0.03$.

The excellent result of this calculation leaves for future research the analytic
determination of the random scaling functions $P_k(\sigma)$. If these could
be written down from first principles the problem of DLA will be settled. In particular
only analytic $P_k(\sigma)$ will allow reliable calculations of $D_q$ for $q\to \infty$
due to the sensitivity to the small $\sigma$ tail where the statistics is low.
\acknowledgments
We thank George Hentschel and Yoram Cohen for helpful discussions and Ellak Somfai
for the numerical code.
This work has been supported in part by the Israel Science Foundation, the US-Israel BSF, and 
the Naftali and Anna
Backenroth-Bronicki Fund for Research in Chaos and Complexity.

\end{document}